\documentstyle[twocolumn,aps]{revtex}
\input psfig
\begin{document}
\bibliographystyle{plain}
\vskip0.5truecm 
\vskip1.truecm

\noindent{\bf Mila and Poilblanc reply:} The main point of Clarke and Strong's 
comment\cite{comment}, namely that 
our results concerning the effect of integrability on the hopping of 
electrons between
chains\cite{mila} are probably not related to the 
notion of coherence proposed in Ref.\cite{clarke}, 
is important and well taken. In fact, we have already stressed in our
work\cite{mila} that
integrability affects the {\it intermediate--time} behaviour but not the 
{\it short--time} one, 
while in Ref.\cite{clarke} the authors make predictions about the long-time
dynamics on the basis of a calculation at short times. The
logical conclusion is then that these two effects are in fact 
different phenomena.
In more recent papers we have numerically investigated other 
physical quantities (spectral function\cite{poilblanc,capponi}, transverse 
conductivity\cite{capponi}) in relation to
the notion of coherence developped in Ref.\cite{clarke}.

The proposal\cite{comment} that the effect seen in Ref.\cite{mila} is related 
to ergodicity is also very plausible. This has also been suggested
as a possible
explanation of the effect of integrability on the conductivity of 1D
systems\cite{castela}. However, as far as we know, 
this remains a conjecture not fully established on firm grounds. 

Now, concerning the details of the discussion of 
Ref.\cite{comment}, we think it is important to
make further clarifications. Clarke and Strong suggest that 
$P(\tau)$, the
probability of finding the system in its initial state at time $\tau$ 
after turning
on the hopping $t_\perp$, is an appropriate test of the notion of coherence of
Ref.\cite{clarke} only for
sufficiently short times, the long--time behaviour of this function 
being related to 
ergodic properties of the system. Our point of view is 
slightly different. 
The notion of 
coherence discussed in Ref.\cite{clarke}
is related to the splitting of the main peaks in the bonding ($k_\perp=0$) and
antibonding ($k_\perp=\pi$)
spectral functions, incoherence meaning that the splitting disappears in the
thermodynamic limit\cite{note}. In fact, the same behaviour can also be 
inferred from a study
of $P(\tau)$ provided that: i) The difference at $\tau=0$ between particle 
numbers on the
two chains $\Delta N$ is equal to 1; ii) A non--symmetrized 
initial wavefunction is used. This
claim is supported by Fig. 1, in which we have compared $P(\tau)$ calculated in 
this way with the results deduced from the spectral 
functions\cite{poilblanc}:
The fundamental frequency of the oscillations is the same, and it is equal to
the splitting between the bonding and the antibonding bands. 
The rapid oscillations that appear in $P(\tau)$ deduced from the spectral
functions are 1D features that are {\it not} related to the problem of
coherence between chains.

We think that these results help clarify the difference between the 
notions of coherence discussed in Ref.\cite{clarke} and Ref.\cite{mila} 
respectively. Let us first define $P(\tau)$ with $\Delta N=1$ 
and let us increase the interactions. A reduction of coherence according to
Ref.\cite{clarke} will
show up in $P(\tau)$ as a {\it shift} of the oscillations toward
larger times, incoherence being achieved when the period becomes infinite. This
is consistent with a decrease of the curvature of $P(\tau)$ at small times 
when coherence disappears. 
Alternatively, one can choose a macroscopic value of $\Delta N$.
When the interactions are switched on, the {\it intensity} of the
oscillations will decrease dramatically unless the underlying model is
integrable, in agreement with the notion of coherence discussed in
Ref.\cite{mila}.
\vskip.4cm

\begin{figure}[hp]
\centerline{\psfig{figure=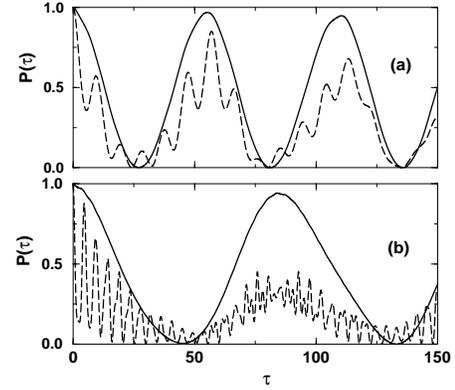,width=6.0cm,angle=0}}
\vspace{0.4cm}
\caption{
Return probability calculated on a $2\times 12$ system at 1/3--filling 
as defined in Ref.\protect\cite{mila} but for 
$\Delta N=1$ (solid line) compared to the
results of Ref.\protect\cite{poilblanc} obtained from the single particle 
transverse Green function (dashed line):
(a) Hubbard model with $U/t=8$ and $t_\perp/t=0.1$; (b) Extended Hubbard model
with $U/t=6$, $V_1/t=3$, $V_2/t=2$ and $t_\perp/t=0.1$.}
\label{fig1}
\end{figure}

In conclusion, we agree that the effect we discussed in Ref.\cite{mila} 
(for $\Delta N>1$) is probably not
related to the notion of coherence introduced in Ref.\cite{clarke}. 
From a numerical point of view, this latter effect is
best studied by looking at physically measurable quantities, like the spectral
function or the transverse optical conductivity. We nevertheless think that 
the function $P(\tau)$ introduced in Ref.\cite{clarke} and used by us in
Ref.\cite{mila} is very useful,
at least pedagogically, in clarifying the difference between the
effects discussed in these papers. 

\vskip.4cm

\noindent Fr\'ed\'eric Mila and Didier Poilblanc

      Laboratoire de Physique Quantique 
      
      Universit\'e Paul Sabatier
      
      31062 Toulouse (France)


\begin{references}
\bibitem{note} The scaling toward the thermodynamic limit is a subtle issue.
The interested reader is urged not to stop at Ref.\protect\cite{poilblanc}, 
but to read Ref.\protect\cite{capponi} as
well.
\end{references}
\end{document}